\documentclass[aps,prb]{revtex4}
\usepackage{amsmath}
\usepackage[russian]{babel}
\usepackage{graphicx}

\begin{document}

\title{Singular resonance in fluctuation-electromagnetic phenomena during the rotation of a nanoparticle near a surface}

\author{A.I. Volokitin$^{*}$}

\affiliation{
Samara State Technical University,  443100 Samara, Russia}

\begin{abstract}
It is shown that   in  fluctuation-electromagnetic phenomena (Casimir force,
Casimir friction,  radiative heat generation) for a spherical nanoparticle with a radius $R$  rotating near a surface a singular
resonance can occur, near which fluctuation-electromagnetic effects are strongly enhanced even
in the presence of dissipation in the system. The resonance takes place at the particle-surface  separation
 $ d <d_0= R[3/4\varepsilon_1''(\omega_1)\varepsilon_2''(\omega_2)]^{1/3}$ (where $\varepsilon_i''(\omega_i)$
is the imaginary part of the dielectric function of
a particle or a medium at the   surface plasmon or phonon polariton frequency  $\omega_i$), when the rotation frequency $\Omega$ coincides with
the poles in the  photon generation rate  at $\Omega\approx \omega_1 + \omega_2$.
These poles arise due to the multiple scattering of electromagnetic waves between the particle and the surface under the conditions of the anomalous Doppler effect and they exist even in the presence of dissipation.
For $ d <d_0$ in the dependence on the particle rotation frequency  the Casimir force can change sign, i.e. the attraction of the particle to the surface is replaced by the repulsion. The obtained results  can be
important for nanotechnology.
\end{abstract}
\maketitle


\vskip 5mm

All material media are surrounded by a fluctuating electromagnetic field because of  thermal and quantum
fluctuations of the current density  inside them. Near the surface this fluctuating electromagnetic field is  strongly enhanced due to  the contribution from evanescent electromagnetic waves. This enhancement is especially large
when on the surface there are surface modes, such as surface plasmon polaritons,
surface phonon polaritons, or vibrational modes of adsorbates. Electromagnetic fluctuations
are the cornerstone of the Casimir physics, which includes   Casimir-Lifshitz-van der
Waals forces  \cite{Casimir1948,Lifshitz1956,Dalvit2011,Klimchitskaya2009,Bimonte2017}, Casimir  friction \cite{Volokitin1999JPCM,Volokitin2008PRB,Volokitin2007RMP,Dedkov2017UPN,Pendry1997JPCM} with its limiting case of quantum friction \cite{Volokitin2007RMP,Pendry1997JPCM,Volokitin2011PRL}, the radiation and the near-field radiative heat transfer \cite{Polder1971,Pendry1999JPCM,Volokitin2001PRB,Volokitin2003JETPLett,Volokitin2004PRB,Joulain2005SSR}.

In recent years considerable progress has been achieved in the study of fluctuation-electromagnetic phenomena. This is due to the development of new experimental methods, which made it possible to investigate these phenomena
on the nanoscale, where they are much stronger than at the microscale. Recently the measurements of the Casimir-Lifshitz-van der Waals forces were  carried out
 with unprecedented accuracy \cite{Dalvit2011,Klimchitskaya2009}. These measurements are consistent with the Lifshitz theory up
to to very large distances, when the retardation effects become important
and when the interaction is determined by thermal, rather than quantum fluctuations. Fluctuation-electromagnetic forces, usually called Casimir forces which in the non-retarded limit are
the van der Waals forces, dominate the interaction between nanostructures and can cause
stiction in small devices such as micro- and nanoelectromechanical systems. As a result of
practical importance of the problem of fluctuation-electromagnetic interaction for the construction of
nano-electromechanical systems and great progress in methods of force detection, experimental
and theoretical studies of the fluctuation-electromagnetic forces between neutral bodies have experienced
an extraordinary rise in the past decade \cite{Volokitiin2018Book,Dalvit2011,Klimchitskaya2009}. A special attention   is devoted to research
of fluctuation-electromagnetic phenomena under dynamic and thermal   nonequilibrium conditions
 \cite{Volokitiin2018Book,Volokitin1999JPCM,Volokitin2008PRB,Volokitin2007RMP,Dedkov2017UPN,Pendry1997JPCM,Polder1971,Pendry1999JPCM,Volokitin2011PRL,
Volokitin2001PRB,Volokitin2004PRB,Joulain2005SSR,Reddy2015AIP,Volokitin2016PRB,Volokitin2016JETPLett,
Volokitin2013EPL,Volokitin2011PRB,Volokitin2001JPCM,Volokitin2017ZNA,Shapiro2017,Podgornik2016PRL,Volokitin2008}.
This interest is due to the fact that in nonequilibrium systems one can
influence on fluctuation-electromagnetic interactions, which is extremely important for designing nano-electromechanical devices.  It was theoretically predicted \cite{Polder1971,Pendry1999JPCM,Volokitin2001PRB,Volokitin2004PRB} and experimentally
 confirmed \cite{Shen2009NanoLett,Greffet2009NatPhotonics,Reddy2015AIP} that the radiative heat flux between two bodies with different temperatures in the
 near field region is many orders of magnitude larger than that determined by the classical Stefan-Boltzmann law.
The relative motion between bodies affects the  Casimir-Lifshitz-van der Waals forces, radiative heat transfer and
leads to  dissipation and Casimir friction \cite{Volokitin1999JPCM,Volokitin2008PRB,Volokitin2007RMP,Dedkov2017UPN,Pendry1997JPCM}, which is one of the mechanisms of non-contact friction.
The results of experiments on the observation of the frictional drag  between quantum wells and graphene sheets, and the current-voltage dependence of graphene on the surface of the polar dielectric SiO$_2$ were explained
using the theory of the Casimir friction \cite{Volokitiin2018Book,Volokitin2011PRL,Volokitin2013EPL,Volokitin2001JPCM,Volokitin2017ZNA}.

At present a great deal of attention is devoted to the study
of rotating nanoparticles in the context of wide variety of
physical, chemical, and biomedical applications. The most
important are related to the use of rotating nanoparticles for
the targeting of cancer cells \cite{Nanomed2014,TrendsBioTech2011,ACSNano2014}. Different
experimental methods for  trapping and rotating nanoparticles were discussed recently in Refs.  \cite{nacommun2011,nanano2013,nanolett2014}. Calculation
of fluctuation-electromagnetic phenomena for two rotating nanoparticles taking into account their mutual polarization due to the multiple scattering of electromagnetic waves using a fluctuation
electrodynamics was done by us in Refs. \cite{Volokitin2017JETPLett,Volokitin2017PRA}. It was shown that a singular resonance is possible for this system which was predicted in Refs. \cite{JacobJOpt2014,JacobOptExp2014,Volokitin2016PRB}. The frictional forces due to quantum fluctuations
acting on a small sphere  rotating near a surface were studied in Refs.  \cite{PendryPRL2012,DedkovEPL2012}  without taking into account the multiple scattering of electromagnetic waves between the particle and the surface.

In this Rapid Communication  a fluctuation electrodynamics is used to calculate the frictional
torque, interaction force, and heat generation for a  nanoparticle rotating near a surface. It is shown that the fluctuation-electromagnetic effects can be greatly enhanced near the singular resonance, which arises from the multiple scattering of electromagnetic waves
between the particle and the surface under the conditions of the anomalous Doppler effect.

\begin{figure}
\includegraphics[width=0.30\textwidth]{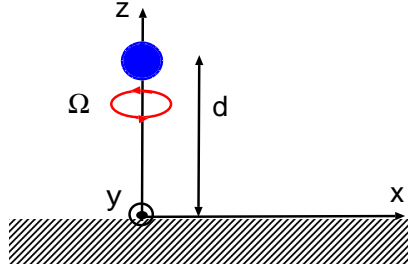}
\caption{A nanoparticle rotating around the $\hat{z}$ axis and located at a distance $d$ from the surface of a semi-infinite
medium.
 \label{Scheme}}
\end{figure}

Consider a spherical particle with a radius $R$ located at a distance $d$ from the surface of a homogeneous
medium which lies in the $xy$ plane. (see Fig. \ref{Scheme}). The particle and the medium have different temperatures  $T_1$ and  $T_2$, and
are characterized by frequency dependent dielectric functions $\varepsilon_1(\omega)$ and $\varepsilon_2(\omega)$, respectively. We introduce
two reference frames $K$ and $K^{\prime}$. In the $K$  reference frame the medium is at rest while the particle rotates around the axis
passing through it   with frequency  $\Omega$. $K^{\prime}$ is the rest reference frame of the particle. The orientation of the rotation axis 
can be arbitrary, but in the present paper we consider the most symmetrical cases when the rotation axis
 is directed along the $\hat{z}$ axis, as in Fig. \ref{Scheme}.  In comparison with the general case in this limiting case
the calculations are much simpler and the obtained results  are qualitatively the same.

According to fluctuation electrodynamics \cite{Volokitiin2018Book,Volokitin2007RMP}, the dipole moment for a polarizable particle  $\mathbf{p}=\mathbf{p}^{f}+\mathbf{p}^{ind}$,
where $\mathbf{p}^{f}$ is the fluctuating dipole moment due to  quantum and thermal fluctuations inside
the particle, $\mathbf{p}^{ind}$ is the induced dipole moment.

In the $K^{\prime}$ reference frame the dipole moment  $\mathbf{p}^{\prime}$ satisfies the equation
\begin{equation}
\mathbf{p}^{\prime}(\omega)=\alpha(\omega)\mathbf{E}^{\prime}(\omega, \mathbf{r}_0) +
\mathbf{p}^{f\prime}(\omega),
\label{dip2perpprime}
\end{equation}
where $\mathbf{E}^{\prime}(\omega, \mathbf{r}_0)$ is the electric field  produced by the fluctuating and induced charge density
of the medium at the position of the particle at $\mathbf{r}=
 \mathbf{r}_0=(0,0,d)$, and the
 polarizability of the particle is determined by the equation
\begin{equation}
\alpha(\omega) = R^3\frac{\varepsilon_1(\omega) -1}{\varepsilon_1(\omega) +2}.
\label{polarizability}
\end{equation}
The relation between the dipole moments of a particle in $ K $ and $K ^{\prime}$ frames is determined by the equations:
$p_{z}^{\prime}(t)=p_{z}(t)$ and
\begin{equation}
\mathbf{p}^{\prime}_{\perp}(t)=
\left(
\begin{array}{cc}
\cos \Omega t& \sin \Omega t\\
-\sin \Omega t &\cos \Omega t
\end{array}\right)\mathbf{p}_{\perp}(t),
\label{T}
\end{equation}
where $\mathbf{p}_{\perp}=(p_x, p_y)$, and for the Fourier components:
$p_{z}^{\prime}(\omega)=
p_{z}(\omega)$,
\begin{equation}
\mathbf{p}^{\prime}_{\perp}
(\omega)=
\hat{e}^{\prime +}p^-({\omega^+}) +
\hat{e}^{\prime -}p^+({\omega^-}),
\label{FTdip}
\end{equation}
where $\omega^{\pm}=\omega \pm \Omega$, $\hat{e}^{\prime \pm}=(\hat{x}^{\prime}\pm
i\hat{y}^{\prime})/\sqrt{2}$, $p^{\pm}=(p_{x}\pm
ip_{y})/\sqrt{2}$. The same relations are valid for  $\mathbf{E}^{\prime}$:
$E_{z}^{\prime}(\omega, \mathbf{r}_0)=
E_{z}(\omega, \mathbf{r}_0)$,
\begin{equation}
\mathbf{E}^{\prime}_{\perp}
(\omega, \mathbf{r}_0)=
\hat{e}^{\prime +}E^-({\omega^+}) +
\hat{e}^{\prime -}E^+({\omega^-}),
\label{FTField}
\end{equation}
where $E^{\pm}=(E_{x}\pm
iE_{y})/\sqrt{2}$, $\mathbf{E}(\omega, \mathbf{r}_0)=\mathbf{E}^f(\omega, \mathbf{r}_0)+\mathbf{E}^{ind}(\omega, \mathbf{r}_0)$, where $ \mathbf{E}^f(\omega, \mathbf{r}_0)$ is the fluctuating electric field due to  quantum and thermal fluctuations inside the medium, $\mathbf{E}^{ind}(\omega)=\mathbf{G}(\mathbf{r}_0, \mathbf{r}_0, \omega)\cdot \mathbf{p}$ is  the induced electric field at the position of the particle, $\mathbf{G}$ is the electromagnetic Green tensor. Near the planar surface  in the electrostatic limit the Green tensor is reduced to  $G_{ij}=G_{ii}\delta_{ij}$ with the components $G_{xx}=G_{yy}=G_{zz}/2=R_p/(2d)^3$, where the reflection amplitude for the surface of the medium
\begin{equation}
R_p(\omega)=\frac{\varepsilon_2(\omega)-1}
{\varepsilon_2(\omega)+1}.
\label{Rp}
\end{equation}
Using these relations in (\ref{dip2perpprime}) and taking into account that
$(\hat{e}^{\pm}\cdot \hat{e}^{\mp})=1$, $(\hat{e}^{\pm}\cdot \hat{e}^{\pm})=0$ and
$(\hat{e}_z\cdot \hat{e}^{\pm})=(\hat{e}^{\pm}\cdot \hat{e}_z)=0$
we get
\begin{equation}
p_{z}(\omega)=\frac{p_{z}^f(\omega) + \alpha(\omega)E_z^f(\omega, \mathbf{r}_0)}
{1-2\alpha(\omega)R_p(\omega)/(2d)^3},
\label{p1z}
\end{equation}
\begin{equation}
E_{z}(\omega, \mathbf{r}_0)=\frac{2R_p(\omega)p_{z}^f(\omega)/(2d)^3 + E_z^f(\omega, \mathbf{r}_0)}
{1-2\alpha(\omega)R_p(\omega)/(2d)^3},
\label{Ez}
\end{equation}
\begin{equation}
p_{x}(\omega)=\frac 1{2}\left[\frac{p^{f\prime+}(\omega^+)
+\alpha(\omega^+)
E^{f +}(\omega)}
{D^+}+\frac{p^{f\prime-}(\omega^-)
+\alpha(\omega^-)
E^{f -}(\omega)}
{D^-}\right],
\label{p2x}
\end{equation}
\begin{equation}
E_{x}(\omega, \mathbf{r}_0)=\frac 1{2}\left[\frac{R_p(\omega)p^{f\prime+}(\omega^+)/(2d)^3
+ E^{f +}(\omega)}
{D^+}+\frac{R_p(\omega)p^{f\prime-}(\omega^-)(2d)^3
+ E^{f -}(\omega)}
{D^-}\right],
\label{Ex}
\end{equation}
\begin{equation}
p_{y}(\omega)=\frac 1{2i}\left[\frac{p^{f\prime+}(\omega^+)
+\alpha(\omega^+)
E^{f +}(\omega)}
{D^+}-\frac{p^{f\prime-}(\omega^-)
+\alpha(\omega^-)
E^{f -}(\omega)}
{D^-}\right],
\label{py}
\end{equation}
\begin{equation}
E_{y}(\omega, \mathbf{r}_0)=\frac 1{2i}\left[\frac{R_p(\omega)p^{f\prime+}(\omega^+)/(2d)^3
+ E^{f +}(\omega)}
{D^+}-\frac{R_p(\omega)p^{f\prime-}(\omega^-)(2d)^3
+ E^{f -}(\omega)}
{D^-}\right],
\label{Ey}
\end{equation}
where $D^{\pm}=1-\alpha(\omega^{\pm})R_p(\omega)/(2d)^3$, $E^{f\pm}(\omega)=
E_{x}^{f}(\omega, \mathbf{r}_0)\pm iE_{y}^{f}(\omega, \mathbf{r}_0)$, $p^{f\pm}(\omega^\pm)=
p_{x}^{f\prime}(\omega^\pm)\pm ip_{y}^{f\prime}(\omega^\pm)$. The spectral density of fluctuations of the dipole moment of a particle in the $K^{\prime}$ rest frame is determined by the fluctuation-dissipation theorem
\begin{equation}
<p^{f}_{i}(\omega)p^{f*}_{j}(\omega^{\prime})>=2\pi \delta(\omega-\omega^{\prime})
<p^{f}_{i}p^{f*}_{j}>_{\omega},
\label{FDT}
\end{equation}
where
\begin{equation}
<p^{f}_{i}p^{f*}_{j}>_{\omega}=\hbar\mathrm{Im}\alpha(\omega)
\mathrm{coth}\left(\frac{\hbar\omega}{2k_BT_1}\right)\delta_{ij},
\end{equation}
and the spectral density of fluctuations of the electric field in the $K$ reference frame  is given by
\begin{equation}
<E^{f}_{i} (\mathbf{r}_0)E^{f*}_{j}(\mathbf{r}_0)>_{\omega}=\hbar\mathrm{Im}G_{ij}(\omega)
\mathrm{coth}\left(\frac{\hbar\omega}{2k_BT_2}\right)\delta_{ij},
\label{FDTE}
\end{equation}
Using (\ref{p1z})-(\ref{FDTE}) for the torque acting on the particle
along the $\hat{z}$ axis we obtain
\[
M_z=\int_{-\infty}^{\infty}\frac{d\omega}{2\pi}
<p_x E_y^*-p_y E_x^*>_{\omega}=
\label{Torque1}
\]

\begin{equation}
=\frac{\hbar}{\pi}\int_{-\infty}^{\infty}d\omega\frac{
\mathrm{Im}\alpha(\omega^-)\mathrm{Im}R(\omega)/ (2d)^3}{|1-\alpha(\omega^-)R_p(\omega)/(2d)^3|^2}
\left(\mathrm{coth}\frac{\hbar \omega}{2k_BT_2}-\mathrm{coth}
\frac{\hbar \omega^-}{2k_BT_1}\right).
\label{Torque2}
\end{equation}
The contribution to the torque from quantum fluctuations (quantum friction), which exists even for
$T_1=T_2=0$ K, is given by the formula
\begin{equation}
M_{zQ}=\frac{2\hbar}{\pi }\int_0^{\Omega}d\omega\frac{
\mathrm{Im}\alpha(\omega^-)\mathrm{Im}R_p(\omega) (2d)^3}{|1-\alpha_1(\omega)\alpha_2(\omega^-)/(2d)^3|^2}
\label{TorqueQ}
\end{equation}
The heat generated in a medium by a fluctuating electromagnetic field is given by
\[
P_2 = -\int_{-\infty}^{\infty}\frac{d\omega}{2\pi}
<\mathbf{j}\cdot \mathbf{E}^*>_{\omega}
=\int_{\infty}^{\infty}\frac{d\omega}{2\pi}
<i\omega\mathbf{p}\cdot \mathbf{E}^*>_{\omega}
\]
\[
 = \frac{\hbar}{\pi (2d)^3}\int_{-\infty}^{\infty}d\omega\omega
\left[\frac{\mathrm{Im}\alpha(\omega)
\mathrm{Im}R_p(\omega)}{|1-2\alpha(\omega)R_p(\omega)/(2d)^3|^2}
\left(\mathrm{coth}\frac{\hbar \omega}{2k_BT_1}-
\mathrm{coth}\frac{\hbar \omega}{2k_BT_2}\right) \right.
\]

\begin{equation}
\left. +\frac{
\mathrm{Im}\alpha(\omega^-)\mathrm{Im}R_p(\omega)}{|1-\alpha(\omega^-)R_p(\omega)/(2d)^3|^2}
\left(\mathrm{coth}\frac{\hbar \omega^-}{2k_BT_1}-\mathrm{coth}
\frac{\hbar \omega}{2k_BT_2}\right)\right].
\label{Q1}
\end{equation}
The heat $P_1$ generated in a particle can be found from the equation: $-M_z\Omega = P_1 + P_2$.

The force acting on a particle along the $\hat{z}$ axis is given by the formula
\[
F_{z} =\int_{-\infty}^{\infty}\frac{d\omega}{2\pi}
<\mathbf{p}\cdot \frac{\partial}{\partial z}\mathbf{E}^*(z\rightarrow d)>_{\omega}
\]
\[
 = -\frac{3\hbar}{8\pi d^7}\int_{-\infty}^{\infty}d\omega
\left[\frac{1}{|1-2\alpha(\omega)R_p
(\omega)/(2d)^3|^2}\left(\mathrm{Im}\alpha(\omega)
\mathrm{Re}R_p(\omega)\mathrm{coth}\frac{\hbar \omega}
{2k_BT_1}+\mathrm{Re}\alpha(\omega)
\mathrm{Im}R_p(\omega)\mathrm{coth}\frac{\hbar \omega}{2k_BT_2}\right)
 \right.
\]
\begin{equation}
\left. +\frac{1}{|1-\alpha(\omega^-)R_p
(\omega)/(2d)^3|^2}
\left(
\mathrm{Im}\alpha(\omega^-)\mathrm{Re}R_p(\omega)\mathrm{coth}\frac{\hbar \omega^-}{2k_BT_1}
+
\mathrm{Re}\alpha(\omega^-)\mathrm{Im}R_p(\omega)\mathrm{coth}
\frac{\hbar \omega}{2k_BT_2}\right)\right]
\label{F1}
\end{equation}
Using the representation of complex quantities in the form $\alpha(\omega)=|\alpha(\omega)|\mathrm{exp}(i\phi_1)$ and $R_p(\omega)=|R_p(\omega)|\mathrm{exp}(i\phi_2)$, for
 $\Omega =0$ the photon tunneling rate   through the vacuum gap between the particle and the medium can be
written in the form
\begin{equation}
t^T=\frac {4\mathrm{Im}R_p(\omega)\mathrm{Im}\alpha(\omega)/(2d)^3}{|1-R_p(\omega)\alpha(\omega)/(2d)^3|^2}=
\frac {4|R_p||\alpha/(2d)^3|\mathrm{sin}\phi_1\mathrm{sin}\phi_2}{1+|R_p|^2|\alpha/(2d)^3|^2-2|R_p||\alpha/(2d)^3|\mathrm{cos}(\phi_1+\phi_2)}.
\label{tT}
\end{equation}
From (\ref{tT}) follows that the tunneling rate is maximal  at $|R_p||\alpha/(2d)^3|=1$ and $\phi_1=\phi_2$ when
$t^T_{max}=1$.
Thus $P\leq P_{max}$, where
\begin{equation}
P_{max}=\frac{\pi k_B^2}{6\hbar}\left(T_2^2 - T_1^2\right).
\end{equation}

The radiative heat transfer   between the particle and the surface is strongly enhanced in the case of
the resonant photon tunneling  \cite{Volokitiin2018Book,Volokitin2007RMP,Volokitin2003JETPLett,Volokitin2004PRB}. The particle and the dielectric surface have resonances at $\varepsilon_1^{\prime}(\omega_1)=-2$  and $\varepsilon_2^{\prime}(\omega_2)=-1$, respectively, where $\varepsilon_i^{\prime}$
 is the real part of $\varepsilon_i$. For a polar dielectric $\omega_i$  determines the frequency
of the surface phonon polariton. Near the resonance at $\omega \approx
\omega_i$ the polarizability of the particle and the reflection amplitude
 for the surface of a dielectric can be written in the form
\begin{equation}
\alpha (\omega) \approx -R^3\frac{a_1}{\omega - \omega_1 +i\Gamma_1},\,\,
R_p(\omega)\approx -\frac{a_2}{\omega - \omega_2 +i\Gamma_2},
\label{res}
\end{equation}
where
\begin{equation}
a_i=\frac {b_i}{(d/d\omega)\varepsilon_i^{\prime}(\omega)|_{\omega=\omega_i}},\,\,\,
\Gamma_i = \frac {\varepsilon_i''(\omega_i)}{(d/d\omega)
\varepsilon_i^{\prime}(\omega)|_{\omega=\omega_i}},
\end{equation}
where $b_1=3$ and $b_2=2$, $\varepsilon_i''$ is the imaginary part of  $\varepsilon_i$.
Near the resonance, which occurs when $\omega_1=\omega_2=\omega_0$, the photon
tunneling rate  at $\Gamma_i\ll a=\sqrt{a_1a_2}$ can be written in the form
\begin{equation}
t^T\approx \frac {4a^2\Gamma_1\Gamma_2(R/2d)^3}{[(\omega-\omega_+)^2+\Gamma^2]
[(\omega-\omega_-)^2+\Gamma^2]}
\label{tapp}
\end{equation}
where $\omega_\pm = \omega_0 \pm a(R/2d)^{3/2} $, $\Gamma = (\Gamma_1+\Gamma_2)/2$. For $a(R/2d)^{3/2}>\Gamma$ the resonant heat transfer is given by the equation
\begin{equation}
P_{res}\approx 3\frac{\hbar \omega_0\Gamma_1\Gamma_2}{\Gamma}[n_1(\omega_0)-n_2(\omega_0)]
\label{Pres}
\end{equation}
where $n_i(\omega)=[\exp(\hbar \omega/k_BT_i)-1]^{-1}$. For $\hbar\omega_0 <
k_BT_i$ we get $P_{res}\approx 3\Gamma k_B(T_2-T_1)$ and for $T_2\gg T_1$
\begin{equation}
\frac{P_{res}}{P_{max}}\approx \frac{18}{\pi}\left(\frac{\hbar \Gamma_1\Gamma_2}{k_BT_2\Gamma}
\right)
\ll \left(\frac{\hbar \omega_0}{k_BT_2}\right)< 1.
\end{equation}
For $a(R/d)^{3/2}<\Gamma$
\begin{equation}
P_{res}\approx 4\frac{\hbar \omega_0 a^2\Gamma_1\Gamma_2}{\Gamma^3}\left(\frac{R}{2d}\right)^3
[n_2(\omega_0)-n_1(\omega_0)]<4\frac{\hbar \omega_0 \Gamma_1\Gamma_2}{\Gamma}[n_2(\omega_0)-n_1(\omega_0)].
\end{equation}

A resonance of a different type is possible for a rotating particle under the conditions of the anomalous Doppler effect, when
 $\omega-\Omega < 0$ \cite{Volokitin2017JETPLett,Volokitin2017PRA}. In this case, instead of the photon tunneling, the photon generation occurs in the particle and the medium.
Taking into account  that in this case $\alpha(\omega-\Omega)=\alpha^*(\Omega-\omega) =|\alpha(\omega-\Omega)|\mathrm{exp}(-i\phi_1)$, the photon emission rate   can be written in the form
\[
t^E=-\frac{4\mathrm{Im}\alpha(\omega-\Omega)
\mathrm{Im}R_p(\omega)/(2d)^3}
{|1-\alpha(\omega-\Omega)R_p(\omega))/2d)^3|^2}=
\]
\begin{equation}
\frac {4|R_p(\omega)||\alpha(\omega-\Omega)/(2d)^3)|\mathrm{sin}\phi_1\mathrm{sin}\phi_2}{1+|R_p(\omega)|^2|\alpha(\omega-\Omega)/(2d)^3)|^2-2|R_p(\omega)||\alpha(\omega-\Omega)/(2d)^3)|
\mathrm{cos}(\phi_1-\phi_2)},
\end{equation}
which diverges ($t^E_{max}=\infty$) for $|R_p(\omega)||\alpha(\omega-\Omega)/(2d)^3|=1$ and $\phi_1=\phi_2$. In this case, the stationary rotation
 is impossible, since the presence of divergence is associated with the electromagnetic instability, when the electromagnetic
 field increases  unrestrictedly  with time due to the conversion  of the rotation mechanical energy into an electromagnetic energy
even if there is dissipation in the system \cite{SilveirinhaNJP2014}. However, near the singular resonance the stationary
rotation  is possible. In this case, the electromagnetic field and the fluctuation-electromagnetic effects will be
strongly enhanced. The product  $|R_p(\omega)||\alpha(\omega-\Omega)/(2d)^3|$ reaches a maximum at $\omega = \omega_2$ and $\Omega=\omega_1+\omega_2$, when
$|R_p(\omega_2)|=
2/\varepsilon_2''(\omega_2)$ and $|\alpha(\omega_1)|=
3R^3/\varepsilon_1''(\omega_1)$. Therefore, the condition for the appearance of a singular resonance has the form
\begin{equation}
\frac{|\alpha(\omega_1)R_p(\omega_2)|}{(2d)^3}=\frac{3}{4\varepsilon_1''(\omega_1)\varepsilon_2''(\omega_2)}\left(\frac {R}{d}\right)^3=1.
\label{resonance}
\end{equation}
From where it follows that a singular resonance arises when $d<d_0$, where
\begin{equation}
d_0=R\left(\frac{3}{4\varepsilon_1''(\omega_1)\varepsilon_2''(\omega_2)}\right)^{1/3}.
\label{rescondition}
\end{equation}
Since the dipole approximation is valid for $R/d\ll 1$,  the appearance of a singular resonance requires the validity of
the condition $\varepsilon_1''(\omega_1)\varepsilon_2''(\omega_2)\ll 1$. For example, the optical properties of silicon carbide (SiC) can be described
using the oscillator model \cite {Palik}
\begin{equation}
\varepsilon (\omega )=\epsilon _\infty \left( 1+\frac{\omega
_L^2-\omega _T^2}{ \omega _T^2-\omega ^2-i\Gamma \omega }\right),
\label{1nine}
\end{equation}
with $\varepsilon _\infty =6.7$, $\omega _L=1.8\cdot 10^{14}$s$^{-1}$, $\omega
_T=1.49\cdot 10^{14}$s$^{-1}$, and $\Gamma =8.9\cdot 10^{11}$s$^{-1}$.
The frequencies of the surface phonon
polaritons for a particle and a semi-infinite medium are determined by the equations $\varepsilon^{\prime}(\omega
_1)=-2$ and $\varepsilon^{\prime}(\omega_2)=-1$.
From where, taking into account (\ref{1nine}), we obtain $\omega _1=1.73\cdot 10^{14}$s$^{-1}$ and $\omega _2=1.76\cdot 10^{14}$s$^{-1}$. The imaginary part of the dielectric  function $\varepsilon''(\omega_1)=0.171$ and $\varepsilon''(\omega_2)=0.137$. Whence for a particle and a medium from silicon carbide the critical  distance $d_0=3.17R$.

For $\omega\approx \omega_2$ the reflection amplitude for the surface is determined by the equation (\ref{res}), and for  $\omega-
\Omega\approx-\omega_1$ the polarizability of a particle can be written in the form
\begin{equation}
\alpha(\omega-\Omega)=\alpha^*(\Omega-\omega)\approx -R^3\frac{a_1}{
\Omega-\omega_2-\omega -i\Gamma_2},
\label{antires}
\end{equation}
Thus, for $\omega \approx \omega_2$ and $\omega-
\Omega\approx-\omega_1$  the photon generation rate for $0<\omega<\Omega$ can be written in the form
\begin{equation}
t^E\approx
\frac{4\Gamma_1\Gamma_2a_1a_2(R/2d)^3}
{(\Gamma_1+\Gamma_2)^2(\omega -\omega_c)^2
+\left[\Gamma_1\Gamma_2\left(\frac{\Omega-\Omega_0}{\Gamma_1+\Gamma_2}
\right)^2
-(\omega -\omega_c)^2+ \frac{(\Omega-\Omega_0)(\Gamma_2-\Gamma_1)(\omega-\omega_c)}
{\Gamma_1+\Gamma_2}+
\Gamma_1\Gamma_2-a_1a_2(R/2d)^3\right]^2}
\label{tantipart}
\end{equation}
where $\Omega_0=\omega_1+\omega_2$,
\begin{equation}
\omega_c=\frac{\Gamma_1(\Omega-\omega_2)+\Gamma_2\omega_1}{\Gamma_1+\Gamma_2}
\end{equation}

For $d<d_0$  the photon generation rate diverges at
 $\omega=\omega_c$ and
$\Omega =
\Omega^{\pm}$, where
\begin{equation}
\Omega^{\pm}=\Omega_0\pm(\Gamma_1+\Gamma_2)\sqrt{\frac{3}{4\varepsilon_1(\omega_1)\varepsilon_2(\omega_2)}
\left(\frac{R}{d}\right)^3-1}.
\end{equation}
Close to the resonance when
\begin{equation}
\frac{\Gamma_1\Gamma_2}{(\Gamma_1+\Gamma_2)^2}\left|\left(\frac{\Omega - \Omega_0}{\Gamma_1+\Gamma_2}\right)^2
+1-\frac{3}{4\varepsilon_1(\omega_1)\varepsilon_2(\omega_2)}
\left(\frac{R}{d}\right)^3\right|\ll 1,
\end{equation}
using (\ref{tantipart}) in (\ref{Q1}), we obtain for the quantum heat generation rate
\begin{equation}
P_{1Q}\approx \frac{2\hbar \omega_c}{\Gamma_1+\Gamma_2}\frac{a_1a_2(R/2d)^3}
{\left|\left(\frac{\Omega - \Omega_0}{\Gamma_1+\Gamma_2}\right)^2
+1-\frac{3}{4\varepsilon_1(\omega_1)\varepsilon_2(\omega_2)}
\left(\frac{R}{d}\right)^3\right|}.
\label{P1Q2}
\end{equation}
For $d> d_0$, the quantum heat generation rate reaches a maximum at $\Omega=\Omega_0$ when for $d\rightarrow d_0$  it diverges as
\begin{equation}
P_{1Q}\propto \frac{d_0}{|d-d_0|}.
\end{equation}

\begin{figure}
\includegraphics[width=0.70\textwidth]{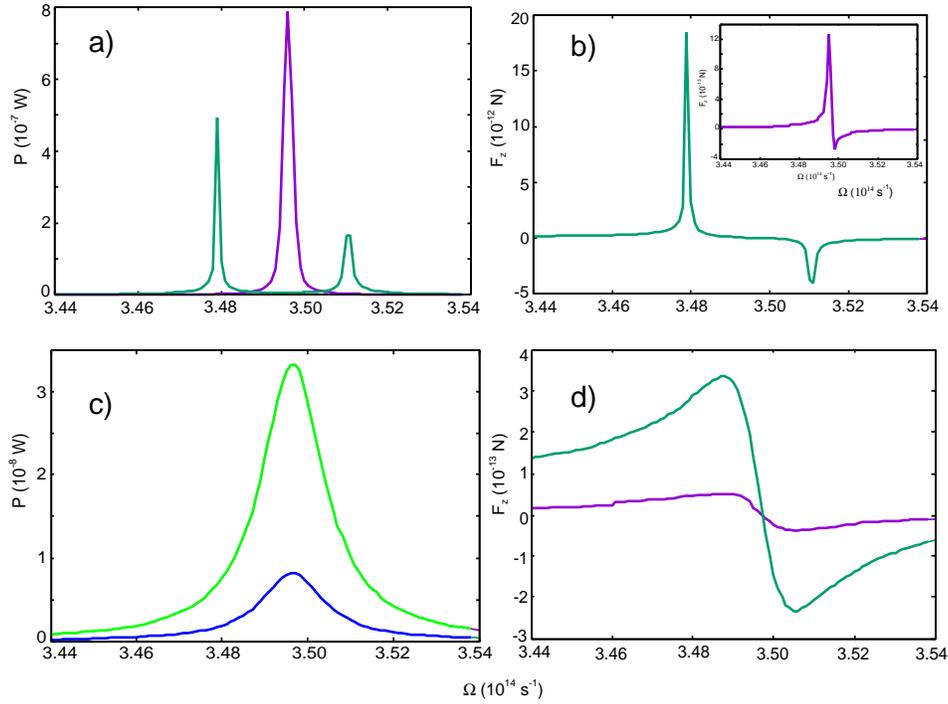}
\caption{a) The dependence of the heat generation rate for a  SiC surface and b) the interaction force between a spherical
SiC particle with a radius $R=0.5$nm and a SiC surface  on the particle rotation frequency $\Omega$. Lines of green and blue
colors show the results of calculations for $d=2R<d_0$ and $d=3.2R>d_0$, respectively, where $d_0=3.17R$ is the critical separation between the particle and the
surface, below which the quantum heat generation rate diverges at the resonant frequencies
$\Omega^{\pm}$; c) and d) are the same as in a) and b), but without taking into account the multiple scattering
of electromagnetic waves between the particle and the surface.
\label{SQ}}
\end{figure}

Fig. \ref{SQ} shows the dependence of a) the heat   generation rate on the SiC surface  and b) the interaction forces
between a SiC particle and a SiC surface on the particle rotation frequency $\Omega$ for $d=2R<d_0=3.17R$ (green lines) and $d=3.2R>d_0$
(blue lines). In accordance with the above theoretical analysis, these dependences have two sharp resonances
for $d <d_0$. For  $d >d_0$ there is only one resonance, which diverges when $d\rightarrow d_0$. For static particles
at $T_2=300$ K and $T_1=0$ K it follows from Eq. (\ref{Pres}) that the resonant photon tunneling gives the contribution to the radiative
heat transfer $P_{res}\approx 10^{-9}$W. In contrast to the static case, for rotating particles the thermal energy generation rate   diverges at the resonance for $d<d_0$ and $\Omega=\Omega^{\pm}$. At a resonance stationary rotation of a particle is impossible, because the frictional force increases unrestrictedly with time. However, stationary rotation is possible near resonance when the heat generation rate  due to the conversion  of the rotation mechanical energy
into the  thermal energy can significantly exceed the static value of the heat generation rate related with the radiative heat transfer. For $d<d_0$, the interaction force in the dependence on the rotation frequency
 changes sign (see Fig. 2b), i.e. the attraction of the particle to the surface is replaced by the repulsion.
Thus, it is possible to tune the interaction force by changing the particle rotation frequency.

\textit{Conclusion}.- Fluctuation electrodynamics was used to calculate the  heat generation,   interaction force and  frictional torque during the rotation  of a nanoparticle  near a surface
 taking into account the multiple scattering of electromagnetic waves between the particle and the surface.
In contrast to the static case, all these quantities diverge at the resonant conditions even in the presence of dissipation in the particle and the medium. The origin of these divergences  are  related with the poles in the photon generation rate
resulting from  the multiple scattering of electromagnetic waves between a particle and a surface under the conditions of the anomalous Doppler effect. The obtained results can find wide application in nanotechnology.
In particular, they can be used to tune the interaction forces and the heat generation by changing
the rotation frequency. These processes can be used for local surface heating and for
nanoelectromechanical systems. When using a rotating nanoparticle, the  diameter
of the surface heating area ($\sim d$) can be much smaller than the diameter of the laser beam. For the practical application of the predicted effects, it is necessary to search for or create materials with a low frequency  of the surface plasmon or phonon polaritons and a small value of the imaginary part of the dielectric function at this frequency.
The polar dielectric SiO$_2$, as well as SiC, has surface phonon polaritons in the IR range, and
InSb semiconductor has the frequencies of surface plasmon-phonon polaritons in the THz band \cite{Palik}. However
for these materials  $\varepsilon''>1$ at the resonant frequencies and  therefore a singular resonance can not arise for them.
A very low frequency of the plasmon polaritons in the GHz region can have metamaterials \cite{Pendry1998}.

\vskip 0.5cm

$^*$alevolokitin@yandex.ru

\end{document}